\documentstyle[aps]{revtex}
\begin{document}
\draft
\preprint{IMSc-99/13; hep-th/9904006}
\title{Does black hole radiance break supersymmetry ?}
\author{Parthasarathi Majumdar\footnote{email: partha@imsc.ernet.in}}
\address{The Institute of Mathematical Sciences, CIT Campus, Madras
600113,  India.}
\maketitle
\begin{abstract}
Arguments presented in an earlier paper, demonstrating the 
breakdown of global supersymmetry in Hawking radiation from a generic four
dimensional black hole with infalling massless scalar and spinor
particles, are reexamined. Careful handling of the Grassmann-valued
spinorial supersymmetry parameter is shown to
lead to a situation wherein supersymmetry may not actually break. A
comparative analysis in flat spacetime at finite temperature is also
presented. 
\end{abstract}

\section{Introduction}

It is now almost universally accepted that any generic black
hole\footnote{Supersymmetric black holes, i.e., those associated with a
covariantly conserved Killing spinor, are usually extremal and do not emit
Hawking radiation. Such special black holes are not considered here.} with
infalling quantum matter fields radiates like a
black body in equilibrium
at a temperature $T_{bh} \sim \kappa/2 \pi$, where, $\kappa$ is the
surface gravity at the horizon of the black hole \cite{hawk}.  
Consequently, infalling bose fields are radiated out in a thermal
Bose-Einstein (Planckian) distribution (modulo some `grey-body' factors)
while fermions are radiated out in a Fermi-Dirac distribution. On the
basis of this alone, one might expect that the supersymmetry manifest at
past
null infinity in a system of bosons and fermions will not survive such
radiation. This expectation seems to be borne out in detail in an
investigation performed two years ago \cite{pm}, in which the standard
criterion of spontaneous supersymmetry breakdown is used, namely the
non-vanishing (or otherwise) of the vacuum expectation value of the
supersymmetry variation
of a fermionic operator at future null infinity. The evaluation of 
the vacuum expectation value (vev) in ref. \cite{pm} follows
Hawking's original approach involving a zero temperature quantum field
theory in the black hole background, and is generic in nature. However,
in that derivation, as elsewhere, one has tended to ignore the
fact that the spinorial supersymmetry parameter is actually an element of
a Grassmann algebra (an $a$-number) rather than a $c$-number. In other
words, unlike a $c$-number parameter which commutes with all
operators of the theory, the supersymmetry parameter {\it anti}commutes
with fermionic operators instead. Clearly, this may have serious
implications for evaluation of Green's functions and the like involving
strings of fermionic and bosonic operators, such as the issue of
supersymmetry breaking entails. Our concern here is with possible
ramifications for black hole radiance. Recall that the phenomenon of
black hole radiance is based upon particle creation in the
gravitational field of a black hole. Thus, evaluation of vevs of operators
defined at future null infinity will
involve matrix elements of such operators between states (at future null
infinity) of non-zero fermion number. It is here that a na\"ive handling
of the supersymmetry parameter is most likely to differ from a careful
one. If sharp disparities arise, the conventional wisdom that black hole
radiance breaks supersymmetry, is bound to be challenged. 

Section II of the paper is a brief recapitulation of the main tenets of
the earlier work. In section III, we attempt a more careful evaluation of
the relevant vacuum expectation value, to see if supersymmetry can indeed
be preserved in black hole radiance. In section IV, a comparative analysis
of the supersymmetric model in flat spacetime in presence of a heat bath
at a finite temperature is presented. Our conclusions and outlook are
presented in the final section. 

\section{The earlier formulation surveyed}

We focus on a situation where, at past null infinity (${\cal I}^-$), there
exists a globally supersymmetric model of noninteracting massless complex
scalar and chiral spinor fields. Now, any state on ${\cal
I}^-$ will evolve into a state on the event horizon (${\cal H}^+$),
belonging to one of two mutually exclusive (in absence of backreaction)
classes, viz., those which are purely outgoing, i.e., have zero Cauchy
data on ${\cal H}^+$ and support on ${\cal I}^+$, and those which have
zero Cauchy data on ${\cal I}^+$ and support on ${\cal H}^+$. As is
well-known \cite{hawk}, an inherent ambiguity in the latter is chiefly
responsible for the thermalization of the radiation received at ${\cal
I}^+$.

The scalar and spinor fields in our model have the following expansion
(at ${\cal I}^-$),
\begin{eqnarray}
{\phi} ~ &=& ~\sum_k \frac {1}{\sqrt{2 \omega_k}}~\left ( a^B_k f^B_k~+~
b_k^{B ~ \dag} 
{\bar f_k^B} \right) ~\nonumber \\
{\psi_+} ~ &=& ~\sum_k \frac {1}{\sqrt{2 \omega_k}}~ \left( a^F_{k,+} 
f^F_k~+~b_{k,-}^{F~ \dag} {\bar f_k^F} \right) u_{k,+}~, \label{flxp}
\end{eqnarray}
where, the $\{ f_k \}$ are complete orthonormal sets of solutions of the
respective field 
equations, with positive frequencies only at ${\cal I}^-$,  and $u_{k,+}$
is a positively 
chiral spinor, reflecting the chirality of 
$\psi_+$. The creation-annihilation operators obey the usual algebra,
with $B$ ($F$) 
signifying Bose (Fermi). The conserved N\"other supersymmetry charge is
given in terms of 
these creation-annihilation operators by (at ${\cal I}^-$)
\begin{equation}
Q_+({\cal I}^-)~=~\sum_k \left( a^F_{k,+} 
{b^B_k}^{\dag}~-~{b^F_{k,-}}^{\dag} a^B_k \right) u_+(k)~,\label{qu} 
\end{equation}
and annihilates the vacuum state $|0_- \rangle$ defined by
\begin{equation}
a^{B,F}_k |0_-\rangle~=~0~=~b^{B,F}_k |0_- \rangle~~. \end{equation} 

The existence of two disjoint classes of states at the horizon, as
mentioned earlier, implies that the fields also admit the expansion
\cite{hawk} 
\begin{eqnarray} 
\phi ~ &=& ~ \sum_k \frac{1}{ \sqrt{2
\omega_k}} ~ \left(~ A_k^B p_k^B~+~B_k^{B~ \dag} {\bar p}_k^B ~+~A_k^{'B}
q_k^B~+~ B_k^{'B~ \dag} {\bar q_k^B} ~ \right)~\nonumber \\ 
\psi_+ ~ &=&
~\sum_k \frac{1}{\sqrt{\omega_k}}~ \left( A_{k,+}^F p_k^F~+~ B_{k,+}^{F~
\dag} {\bar p}_k^B ~+~A_{k,+}^{'F} q_k^F~+~ B_{k,-}^{'F~ \dag} {\bar
q_k^F} \right)~u_+(k)~~, \label{flxp+} 
\end{eqnarray} 
where, $\{p_k\}$ are
purely outgoing orthonormal sets of solutions of the respective field
equations with positive frequencies at ${\cal I}^+$, while $\{q_k\}$ are
orthonormal sets of solutions with no outgoing component. The final vacuum
state $|0_+ \rangle$, defined by the requirement 
\begin{equation} 
A^{B,F}|0_+ \rangle~=~0~=B^{B,F} |0_+ \rangle ~=~A'^{B,F}
|0_+ \rangle~=~B'^{B,F} |0_+ \rangle~
\end{equation}
is not unique, because of the inherent ambiguity in defining positive
frequency for the $\{q_k\}$; in fact, one can write $|0_+\rangle = |0_I
\rangle |0_H \rangle$
with the unprimed (primed) operators acting on $|0_I>$ ($|0_H>$). A 
supersymmetry charge
$Q({\cal I}^+)$ may indeed be defined, analogously to eqn. (\ref{qu}), in
terms of the unprimed operators, and that $Q({\cal I}^+) |0_+
\rangle~=~0$. Such
a charge also satisfies the $N=1$ superalgebra at ${\cal I}^+$.

The field operators $a_k, b_k$ at ${\cal I}^-$ are of course related to the $A_k, B_k$ 
and $A'_k, B'_k$ through the Bogoliubov transformations
\begin{eqnarray}
A_k^B~&=&~ \sum_{k'} \left(~\alpha_{kk'}^B a_{k'}^B~+~\beta_{kk'}^B 
b_{k'}^{B ~\dag}~\right)~\nonumber\\
B_k^B~& =&~ \sum_{k'} \left(~\alpha_{kk'}^B b_{k'}^B ~+~\beta_{kk'}^B
a_{k'}^{B~\dag}~ 
\right)~\nonumber \\
A_{k,+}^ F~&=&~\sum_{k'} \left(~\alpha^F_{kk'} a^F_{k', +}~+~\beta^F_{kk'}
b_{k', 
-}^{F~ \dag}~\right)~ \nonumber \\
B^F_{k',- }~&=&~\sum_{k'} \left(~\alpha_{kk'}^F b_{k',-}^F~+~\beta_{kk'}^F
a^{F~ \dag}_{k', 
+}~\right)~, \label{bog}
\end{eqnarray}
and similarly for the primed operators. We notice in passing that  
\begin{equation}
Q({\cal I}^-) |0_+ \rangle~\neq~ 0~,~ Q({\cal  I}^+) |0_-
\rangle~\neq~0~,~ for 
~\beta^{B,F}~\neq~0~.\label{hnt} \end{equation} 

The issue that we now wish to focus on is whether the radiated system of 
particles has $N=1$ spacetime supersymmetry. To address this question,
recall that vacuum 
expectation values (vevs) of observables at future null infinity are
defined by \cite{hawk}
\begin{equation}
\langle~ {\cal O}~ \rangle~\equiv~\langle 0_-|~ {\cal O}~ |0_-
\rangle~=~Tr~
(\rho ~{\cal O})~ \label{vev}
\end{equation}
where, $\rho$ is the density operator. The trace essentially averages over the 
(nonunique) states going through the horizon, thus rendering the vevs of
observables (at 
${\cal I}^+$) free of ambiguities. We also recall that a 
sufficient condition for spontaneous supersymmetry breaking is the
existence of {\it a} 
fermionic operator ${\cal O}$ which, upon a supertransformation, yields
an operator with 
non-vanishing vev, i.e., $ \langle~ \delta_S {\cal O}~ \rangle~\neq~0$.
Thus, if one is
able to show that 
for {\it all} fermionic observables ${\cal O}({\cal I}^+)$,
\begin{equation}
\langle~ \delta_S {\cal O}({\cal I}^+)~ \rangle~=~0~, \label{ss}
\end{equation}
then we are guaranteed that the outgoing particles form a supermultiplet. 

However, this is not the case, as is not difficult to see; for, consider
the supercharge 
operator itself at ${\cal I}^+$. Using the supersymmetry algebra, it can
be shown that
\begin{equation}
\langle~ \delta_S {\bar Q}_{\dot \alpha}({\cal I}^+)~
\rangle~=~\epsilon^{\beta}~ \langle~ P_{\beta {\dot \alpha}} ({\cal 
I}^+)~ \rangle~,\label{mom} \end{equation}
where $P_{\beta {\dot \alpha}}$ is the momentum operator of the theory
and $\epsilon^{\beta}$ the spinorial supersymmetry parameter. 
In our free field theory, the rhs 
of (\ref{mom}) is trivial to calculate, using eq. (\ref{bog}) above, so
that we obtain, suppressing obvious indices, 
\begin{eqnarray}
\langle~\delta_S {\bar Q}({\cal I}^+) ~\rangle~ &=& \epsilon \sum_k k ~
\langle~N^B_k~+~N^F_k~ \rangle~\nonumber \\
&=& \epsilon \sum_{k,k'} k \left(~ 
|\beta_{kk'}^B|^2~+~|\beta^F_{kk'}|^2 ~\right)~~.\label{ssbr} 
\end{eqnarray}
Thus, supersymmetry is seemingly spontaneously broken in the sense
described above, so long as the Bogoliubov coefficients $\beta^{B,F}$ are
non-vanishing In fact, we know from Hawking's seminal work \cite{hawk}
that 
\begin{eqnarray}
\langle ~N_k^B~ \rangle~ &=& ~ \sum_{k'}
|\beta^B_{kk'}|^2~=~|t^B_k|^2~\left(~e^{2\pi 
|k|/\kappa}~-~1~\right)^{-1}~\nonumber \\
\langle~N_k^F~\rangle~ &=&~ \sum_{k'}
|\beta^F_{kk'}|^2~=~|t^F_k|^2~\left(~e^{2\pi|k|/ 
\kappa}~+~1~\right)^{-1}~.\label{haw}
\end{eqnarray}
Here, $\kappa$ is the surface gravity of the black hole, and
$|t_k^{B,F}|^2$ the transmission coefficients through the potential barrier of the
black hole for bosons, fermions respectively. 

\section{A more careful evaluation}

Using the definition (\ref{vev}) above, the aim is
to evaluate the trace $Tr~\left(~ \rho~[~\epsilon~ Q~,~ {\bar
Q}~]~\right) $. Treating $\epsilon~Q $ as a {\it bosonic} operator and
using the cyclicity of traces, it is easy to see that,
\begin{equation}
\langle~ \delta_S {\cal O}~\rangle~=~Tr~\left(~ [~{\bar
Q}~,~\rho~]~\epsilon~Q~\right)~. \label{trc}
\end{equation}
Thus, the issue of supersymmetry breakdown now depends crucially on
whether the density operator $\rho$ commutes with the supersymmetry
generator ${\bar Q}$ at ${\cal I}^+$. As an operator relation this is
not obvious since we do not know the density operator as a function of
the basic field operators $A^{B,F}~,~B^{B,F}$. In Hawking's approach,
one can only unambiguously determine the diagonal elements of the
density matrix. One would expect the determination of such elements to
be enough to ascertain whether the vev $\langle~\delta_S {\bar
Q}~\rangle$ is non-vanishing.

The basic point of departure from earlier approaches (including ours) is
the property that for any fermionic
operator ${\cal O}~,~ \epsilon {\cal O} = - {\cal O} \epsilon$. That is
to say that $\{ \epsilon, A^F \} = 0$ and similarly for $B^F$. It
follows that, for normalized states (at ${\cal I}^+$) with $n^F_k$
fermions with momentum $k$, we must have
\begin{equation}
\langle n^F_k|~ \epsilon~ | n^F_k \rangle ~=~(-)^{n^F_k} \epsilon~. 
\label{enef}
\end{equation}
In our earlier approach \cite{pm}, the {\it rhs} of eq. (\ref{enef}) 
would not have contained the first factor. This does have an immediate
import for our calculation of $\langle~\delta_S {\cal O}~\rangle $ above
in eq. (\ref{mom}). Rather than expanding the commutator in the
variation $\delta_S {\bar Q}$ as done above, we follow our earlier
step eq. (\ref{vev}) of using the supersymmetry algebra and rewriting
(\ref{vev}) as given in eq. (\ref{mom}),
\begin{equation}
\langle~ \delta_S {\bar Q}_{\dot \alpha}({\cal I}^+)~
\rangle~=~Tr ~\left(~\epsilon^{\beta}~ ~ P_{\beta {\dot \alpha}} ({\cal 
I}^+)~ \rho~\right ).\label{mom1} 
\end{equation}

The Hilbert space of this non-interacting theory ${\cal H} \sim {\cal H}_B
\otimes {\cal H}_F$ so that, changing to the occupation number basis of
the infinite system of uncoupled bose and fermi oscillators, labelled by
momentum $k_{\alpha {\dot \beta}}$, the states are expressed as
$|n^B_k~,~n^F_k~\rangle $. Assuming without loss of generality, a discrete
momentum spectrum, eq. (\ref{mom1}) may be expressed as 
\begin{equation}
\langle~ \delta_S {\bar Q}_{\dot \alpha}({\cal I}^+)~
\rangle~=~\epsilon^{\beta}~\sum_k k_{\beta
{\dot \alpha}}~ \sum_{n^B_k, n^F_k}~\rho_{n^B_k n^F_k,n^B_k
n^F_k}~(-)^{n^F_k}~\left(~n^B_k~+~n^F_k~\right)~. \label{rho}
\end{equation}
Since the relevant density matrix elements are
uniquely determined by states at future infinity (the horizon states being
averaged over) where the system of particles are still non-interacting,
the density operator can be factorized as $\rho = \rho^B ~\rho^F$ where
$\rho^B (\rho^F)$ acts on bosonic (fermionic) states $|n^B_k \rangle$
($|n^F_k \rangle $) alone respectively. 

Recall now the elementary fact that for a given momentum $k$,
$n^F_k~=~0~,~1$. Using eq. (\ref{enef}), (\ref{mom1}) immediately yields
\begin{equation}
\langle~ \delta_S {\bar Q}_{\dot \alpha}~\rangle~=~\epsilon^{\beta}~
\sum_k k_{\beta
{\dot \alpha}}~ \sum_{n^B_k=0}^{\infty}~n^B_k~\left(
~\rho^B_{n^B_k,n^B_k}~\rho^F_{00}~-~\rho^B_{n^B_k-1,n^B_k-1}~\rho^F_{11}  
~\right)~~. \label{dens}
\end{equation}
Not surprisingly, bose-fermi pairing, characteristic of supersymmetric
theories, seems to appear here as well. Thus, if the density operator does
indeed commute with the supersymmetry generator at future asymtopia, the
{\it rhs} of eq. (\ref{dens}) would vanish, implying unbroken
supersymmetry despite a thermal spectrum of radiated particles. However,
even in the explicit form (\ref{dens}), it is not obvious how this
happens. Using eq. (\ref{haw}) and standard properties of density matrices, one
can compute
the {\it rhs} of (\ref{dens}) and obtain,
\begin{eqnarray}
\langle~ \delta_S {\bar Q}_{\dot \alpha}~\rangle~&=&~\epsilon^{\beta}~
\sum_k k_{\beta {\dot \alpha}}~(e^{4\pi |k|/ \kappa}~-~1)^{-1}~[~ (|t^B_k|^2~ +~
|t^F_k|^2)~\nonumber \\
&+&~ e^{2\pi |k|/ \kappa}~(|t^B_k|^2~-~ |t^F_k|^2) -2 |t^B_k|^2~|t^F_k|^2~ ]~.
\label{res} 
\end{eqnarray}
Thus, whether the {\it rhs} of (\ref{res}) vanishes or not, depends on the nature
of the transmission coefficients (grey body factors) $|t^{B,F}_k|^2$. The
calculation of these coefficients depends on explicit solutions of the matter
field equations in specific
black hole backgrounds - a task we do not attempt here. However, we note that in the
limit $|t^B_k|^2 = |t^F_k|^2 = 1$, i.e., in the limit that the
potential barrier of the black hole for outgoing particles is {\it strictly}
reflectionless, $\langle~\delta_S {\bar Q}~\rangle~=~0$. Alternatively, the {\it
lhs} of eq. (\ref{res}) vanishes whenever,
\begin{equation}
|t^F_k|^2~~=~~{|t^B_k|^2~(e^{2\pi |k|/ \kappa} + 1) \over {(e^{2\pi |k| / \kappa}
-1) +2|t^B_k|^2 }}~. \label{nul} 
\end{equation}
Of course, both the above requirements are non-generic: the first, namely that the
potential barrier is reflectionless, is extremely unrealistic. As for the second,
namely eq. (\ref{nul}), this is something which only detailed calculation of grey
body factors $|t^{B,F}_k|^2$ for specific black hole metrics can verify.  
Observe, using eq. (\ref{haw}) that eq. (\ref{nul}) can be rewritten as 
\begin{equation}
\langle~N_k^F~\rangle~~=~~{\langle~N_k^B~\rangle \over
(1~+~2\langle~N_k^B~\rangle)} ~. \label{ssrel}
\end{equation}
 
\section{Supersymmetry in Minkowski space at Finite Temperature}

A comparison, with the behaviour of the system of massless supersymmetric
bosons and fermions in flat spacetime at a finite temperature
$\beta^{-1}$, is in order. In this case, we have full
knowledge of the density operator as a function of the basic field
operators \cite{fey}, 
\begin{equation} 
\rho~~=~~e^{-\beta~H}~/Tr~e^{-\beta H}~~, \label{temp}
\end{equation}
where, the Hamiltonian $H~=~H^B~(A^B,B^B)~+~H^F(A^F,B^F)$. Using eq. (\ref{trc}),
and recalling that there is no {\it explicit} supersymmetry breaking at ${\cal
I}^+$ so that $[{\bar Q}~,~e^{-\beta H}] =0$, we obtain, 
\begin{equation}
\langle~ \delta_S {\bar Q}_{\dot \alpha}~\rangle~~=0~~.
\end{equation}

More explicitly, taking recourse once again to (\ref{enef}), one can actually
calculate the required vev: it is straightforward to see that the thermal average 
\begin{equation}
\langle~\delta_S {\bar 
Q}~\rangle_{\beta}~=~Tr~\{~\epsilon~H~e^{-\beta~H} \}
/Tr~e^{-\beta~H}~~. \label{thav} \end{equation}
Now, 
\begin{equation}
Tr~\{~\epsilon~H~e^{-\beta~H} \}~=~-{d \over d
\beta}~Tr~\epsilon~e^{-\beta~H}~~. \label{der} \end{equation}
Therefore, to obtain preservation of supersymmetry at a non-zero
temperature, all we have to prove is that $Tr~\epsilon~e^{-\beta~H}$ is
independent of $\beta$.

To see this, we use the fact that the Hamiltonian $H$ is actually a sum of
an infinite number of bosonic and fermionic (spin 1/2) harmonic
oscillator Hamiltonians, each at a frequency $\omega_k~=~|k|$. Thus,
\begin{eqnarray}
Tr~\epsilon~e^{-\beta~H}~&=&~~\sum_k~\sum_{n^B_k~,~n^F_k}~\langle 
n^B_k,n^F_k|~  \epsilon~e^{-\beta~H}~|n^B_k,n^F_k \rangle~ \nonumber \\
&=&~~\epsilon \sum_{k;n^B_k, n^F_k}~(-)^{n^F_k}~\exp
\{-\beta~\omega_k~(~n^B_k~+~n^F_k~)~\}~~\nonumber \\
&=&~~\sum_k~\sum_{n^B_k=0}^{\infty}~
\left(~e^{-\beta~n^B_k~\omega_k}~-~e^{-\beta~(n^B_k~+~1)~\omega_k}~\right)~~.
\label{qed}
\end{eqnarray}
It is clear that there is a term by term cancellation for each value of
$n^B_k$ in the {\it rhs} of eq. (\ref{qed}), {\it except} for the first
term for $n^B_k=0$, which of course is {independent of $\beta$}. Thus, {\it just
because bosons and fermions obey different statistics at a finite
temperature, it is hasty to conclude that supersymmetry is broken.} This
fact was first pointed out by L. van Hove \cite{vh} and subsequently by
other workers \cite{par}. Our formulation here makes only
implicit use of the Klein operator $(-)^{\hat {N^F}}$ in contrast to its 
explicit use in those earlier papers.

\section{Conclusions}

The conditions which lead to $\langle~\delta_S {\bar Q}~\rangle~=~0$, namely that
either the potential barrier of the black hole is reflectionless, {\it or} eq.
(\ref{nul}) above holds, are both non-generic,
requiring calculation of grey body factors for specific black holes. Unlike in
section II where, by
regarding the supersymmetry parameter as a $c-number$ and blithely moving it
outside of vevs, we were able to show generically that supersymmetry is broken,
now it seems that the situation is actually more complicated. The comparison
with the flat space finite temperature case in section IV underlines this
feature fairly well: if none of the conditions for $\langle~\delta_S {\bar
Q}~\rangle~=~0$ hold, then the results do not guarantee that a black body
spectrum is all there is to black hole radiation. The disparity with the
flat space behaviour needs to be probed more thoroughly. We hope to report
on this in the near future. 

It is important to point out that even if one suceeded in 
establishing $\langle~\delta_S {\bar Q}~\rangle~=~0$, in principle there
could be other fermionic operators whose supersymmetry variations have
nonvanishing vevs. However, the simplicity of the model under
consideration makes it very unlikely that the foregoing will be challenged
in any manner. It is of course quite another question if interacting
fields are considered. The calculation of the vev will then have to done
perturbatively, in general, as one would analyze the flat space $T \neq 0$
situation. However, the latter situation has already been considered in
\cite{par}: appropriate use of the properties of the supersymmetry
parameter seems to preserve all supersymmetry Ward identities. So our
result should go through in that case as well.

 \vglue .3in

I thank R. Kaul and A. Dasgupta for useful discussions.

\end{document}